\documentclass[sigconf]{acmart}
 \usepackage{amsmath,amssymb,amsfonts,bm}
\usepackage{graphicx}
\usepackage{textcomp}
\usepackage{xcolor}
\usepackage{float}
\usepackage{array}
\usepackage{tabularx}
\usepackage{array}
\usepackage{balance}
\usepackage{marvosym}
\usepackage{multirow}
\usepackage{array}
\usepackage{enumitem}
\usepackage{booktabs}
\usepackage{algpseudocode}
\usepackage[ruled]{algorithm2e} 
\usepackage{xcolor}
\usepackage{pifont}
\usepackage{url}

\AtBeginDocument{%
  \providecommand\BibTeX{{%
    \normalfont B\kern-0.5em{\scshape i\kern-0.25em b}\kern-0.8em\TeX}}}

\copyrightyear{2021} 
\acmYear{2021} 
\acmDOI{10.1145/3459637.3482426}
\setcopyright{acmcopyright}\acmConference[CIKM '21]{Proceedings of the 30th ACM International Conference on Information and Knowledge Management}{November 1--5, 2021}{Virtual Event, QLD, Australia}
\acmBooktitle{Proceedings of the 30th ACM International Conference on Information and Knowledge Management (CIKM '21), November 1--5, 2021, Virtual Event, QLD, Australia}
\acmPrice{15.00}
\acmISBN{978-1-4503-8446-9/21/11}

\begin{document}
\fancyhead{}
\title{Double-Scale Self-Supervised Hypergraph Learning for\\ Group Recommendation}

\author{Junwei Zhang}
\orcid{}
\affiliation{%
  \institution{Chongqing University}
  \streetaddress{}
  \city{}
  \state{}
  \country{}
  \postcode{}
}
\email{jw.zhang@cqu.edu.cn}
  
\author{Min Gao}
\authornote{Min Gao is the corresponding author.}
\affiliation{%
  \institution{Chongqing University}
  \streetaddress{}
  \city{}
  \country{}}
\email{gaomin@cqu.edu.cn}

\author{Junliang Yu}
\affiliation{%
  \institution{The University of Queensland}
  \city{}
  \country{}
}
\email{jl.yu@uq.edu.au}

\author{Lei Guo}
\affiliation{%
 \institution{Shandong Normal University}
 \streetaddress{}
 \city{}
 \state{}
 \country{}}
 \email{leiguo.cs@gmail.com}

\author{Jundong Li}
\affiliation{%
  \institution{University of Virginia}
  \streetaddress{}
  \city{}
  \state{}
  \country{}}
\email{jl6qk@virginia.edu}

\author{Hongzhi Yin}
\affiliation{%
  \institution{The University of Queensland}
  \streetaddress{}
  \city{}
  \state{}
  \country{}
  \postcode{}}
\email{h.yin1@uq.edu.au}


\begin{abstract}
With the prevalence of social media, there has recently been a proliferation of recommenders that shift their focus from individual modeling to group recommendation. Since the group preference is a mixture of various predilections from group members, the fundamental challenge of group recommendation is to model the correlations among members. Existing methods mostly adopt heuristic or attention-based preference aggregation strategies to synthesize group preferences. However, these models mainly focus on the pairwise connections of users and ignore the complex high-order interactions within and beyond groups. Besides, group recommendation suffers seriously from the problem of data sparsity due to severely sparse group-item interactions. In this paper, we propose a self-supervised hypergraph learning framework for group recommendation to achieve two goals: (1) capturing the intra- and inter-group interactions among users; (2) alleviating the data sparsity issue with the raw data itself. Technically, for (1), a hierarchical hypergraph convolutional network based on the user- and group-level hypergraphs is developed to model the complex tuplewise correlations among users within and beyond groups. For (2), we design a double-scale node dropout strategy to create self-supervision signals that can regularize user representations with different granularities against the sparsity issue. The experimental analysis on multiple benchmark datasets demonstrates the superiority of the proposed model and also elucidates the rationality of the hypergraph modeling and the double-scale self-supervision. 
\end{abstract}




\maketitle

\section{Introduction}
\begin{figure}[th]
\includegraphics[width=0.4\textwidth]{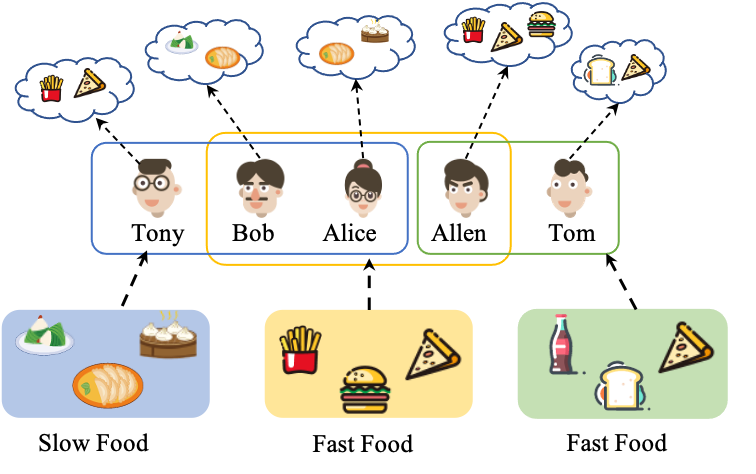}
\caption{An illustrative example of group recommendation.}
\label{f1}
\vspace{-10px}
\end{figure}

Owing to the rapid development of social media in recent years, users with similar interests now have opportunities to form various online groups \cite{DBLP:series/sbcs/YinC16, DBLP:conf/aaai/SunQCLNY20}. Traditional recommender systems designed for individuals can no longer serve the interest of groups. There is an urgent need to develop practical group recommender systems \cite{DBLP:conf/kdd/YuanCL14, DBLP:conf/recsys/BaltrunasMR10, DBLP:conf/sigir/Cao0MAYH18}. The group recommendation aims to reach a consensus among group members, generating suggestions that can cater to most group members \cite{DBLP:conf/aaai/HuCXCGC14, DBLP:journals/jois/MeenaM20}. However, the influences of users vary from group to group, leading to hard decision-making. An illustrative example is shown in Figure~\ref{f1}: Alice and her colleagues (Bob and Tony) form a group that are fond of slow food. However, when Alice and her boss Allen (who likes fast food) go for lunch, fast food would be the first choice due to Allen's higher position. Meanwhile, since Allen has an appetite for pizza, pizza is also recommended to Alice and Bob. It is natural that collective decisions tend to be dynamic, i.e., a group's preference may vary due to the influence of other user members and other groups' preferences. As such, a critical issue is to model the complex interactions of users and groups in the research of group recommendations.

However, existing group recommendation models fail to model the interactions of users and groups well. Many of them adopt heuristic rules, such as average \cite{DBLP:conf/recsys/BaltrunasMR10, DBLP:conf/recsys/BerkovskyF10}, least misery \cite{DBLP:journals/pvldb/Amer-YahiaRCDY09}, and maximum satisfaction \cite{DBLP:series/sci/BorattoC11}, which ignore the impact of user interactions. With the success of deep learning, several group recommendation models 
employ the attention mechanism over user-item graphs to exploit user interactions and synthesize group preferences \cite{DBLP:conf/icde/YinW0LYZ19, DBLP:conf/sigir/TranPTLCL19, DBLP:conf/sigir/HeCZ20}. Nevertheless, these models keep attention on the pairwise connections of users and ignore the complex high-order interactions within and beyond groups. Besides, since many groups are formed temporally, the group-item interaction records are very sparse, making it more difficult to learn an accurate group preference directly from the interaction data \cite{DBLP:journals/corr/abs-2010-00813, DBLP:conf/sigir/HeCZ20, DBLP:conf/icde/YinW0LYZ19}. To tackle the aforementioned two limitations, in this paper, we propose a self-supervised hypergraph learning framework for group recommendation to achieve two goals: (1) capturing the intra- and inter-group interactions among users; (2) alleviating the data sparsity issue with the raw data itself. 
To handle the interaction issue, we explicitly model the complex tuplewise relationships as a hierarchical hypergraph, which aggregates the correlated users within and beyond groups. Distinct from the edge in simple graphs, the hyperedge can connect multiple nodes, which is a natural way to capture tuplewise relations. The proposed hierarchical hypergraph consists of two levels: user- and group-level, where each level of hypergraphs contains the corresponding type of nodes. In the user-level hypergraph, each hyperedge connects the set of users belonging to the same group. To capture the intra-group interactions among different users, we adopt the hypergraph convolutional network to aggregate the related users and generate the dynamic user embeddings. Meanwhile, for exploiting the informative inter-group interactions among users, we adopt triangular motifs \cite{DBLP:conf/aaai/ZhouYR0Z18} which allow each group to select the most relevant groups to acquire information to form the hyperedges in the group-level hypergraph.

Despite the benefits of hypergraph modeling, the group recommendation performance is still compromised by the data sparsity problem, which not only exists in group interaction but also in user interaction. To alleviate this issue, we integrate self-supervised learning (SSL), which can discover self-supervision signals from the raw data, into the training of the hierarchical hypergraph convolutional network. Many existing SSL-based models commonly conduct node/edge dropout in the user-item graph to augment supervised signals \cite{DBLP:conf/nips/YouCSCWS20, DBLP:conf/kdd/QiuCDZYDWT20, DBLP:conf/icml/HassaniA20}. However, this may drop some essential information, especially when the raw group is very sparse, leading to unserviceable supervision signals. Therefore, we propose a double-scale node dropout strategy to create supervisory labels with different and finer granularities. The intuition behind is that not all user-user interactions significantly affect the final group decision, and leveraging partial information can help learn invariant representations. Meanwhile, two different-grained views may to some degree prevent the loss of essential information (e.g. drop important nodes). Concretely, at a coarse granularity, we randomly remove some nodes from the user-level hypergraph structure. As a result, when a removed node belongs to multiple groups, it will disappear in all related hyperedges. As for the fine-grained node dropout, we remove some member nodes only from a specific group, which does not affect this node's existence in other groups. We then maximize the mutual information between the node representations learned from two granularities to regularize the user and group representation against the sparsity issue. Finally, we unify the recommendation task and the self-supervised learning task to optimize model parameters by a two-stage training approach. 
\par
In summary, the main contributions of this paper are as follows:

\begin{itemize}[leftmargin=*]
\item We devise a hierarchical hypergraph learning framework to capture the intra- and inter-group interactions among users in the group recommendation. 
\item We propose a SSL strategy with different granularities to enhance user and group representations and alleviate the data sparsity problem, which is seamlessly coupled with the hierarchical design of the hypergraph convolutional network.
\item We conduct extensive experiments on three group recommendation datasets to exhibit the superiority of the proposed model over some recent baselines and elucidate why the hypergraph learning and the double-scale SSL strategy can improve group recommendation. The code implementation of our model is released at \url{https://github.com/0411tony/HHGR}.
\end{itemize}

 The remainder of this paper is organized as follows. We first provide preliminaries of the hypergraph and the definition of the group recommendation in Section 2. Then, we present our proposed model in detail in Section 3. Section 4 describes experimental research. Next, we further discuss the related work in Section 5. Finally, we summarize our work and look forward to future work in Section 6.

\begin{figure*}[ht]
\includegraphics[width=0.85\linewidth]{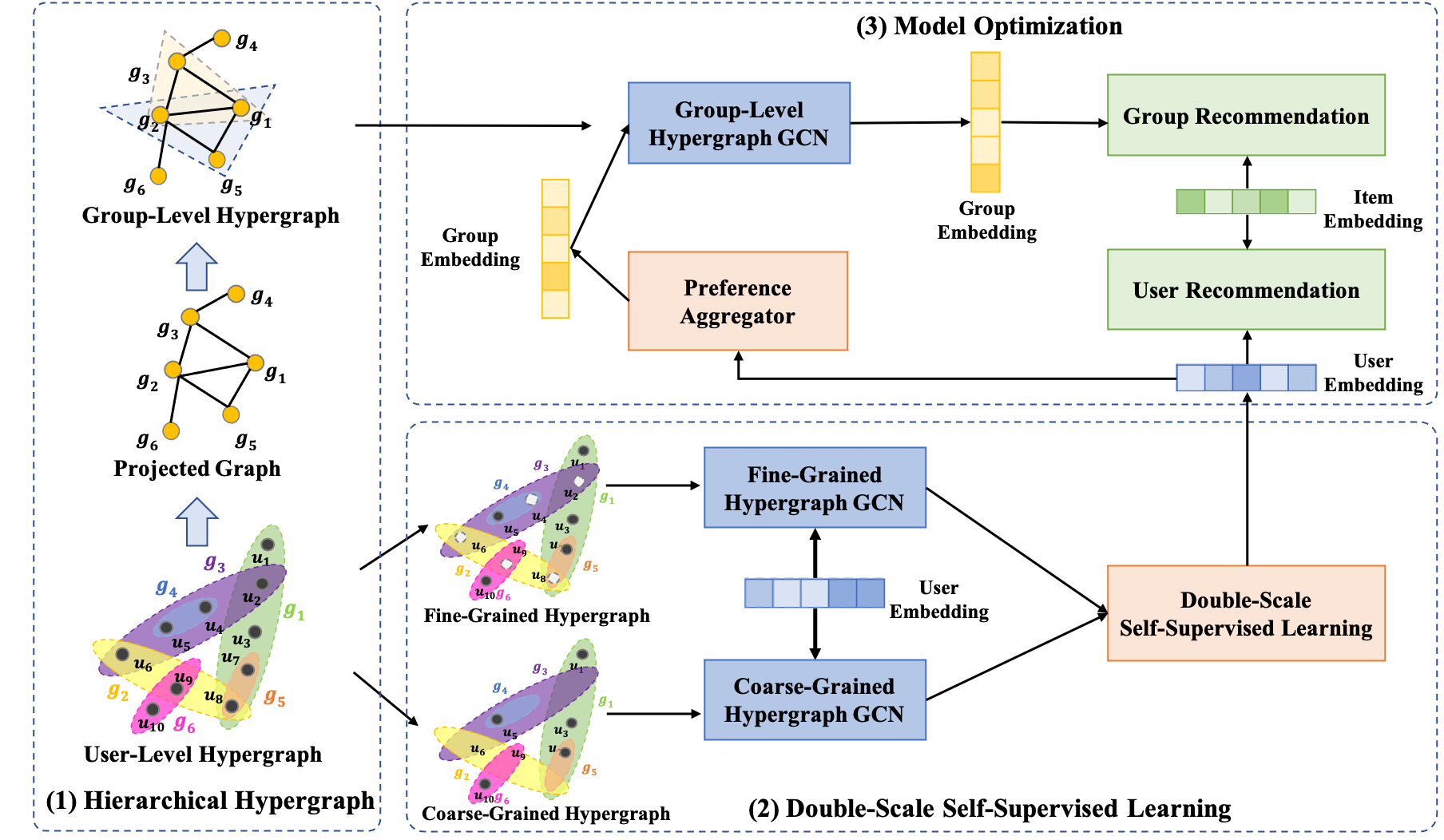}
\caption{The overview of our proposed framework for group recommendation.}
\label{f2}
\end{figure*}

\section{Preliminaries}

To facilitate understanding, we present the definition of the hypergraph and the task of the group recommendation in this section.

\subsection{Definition of Hypergraph}
The hypergraph is defined as $G=(\mathcal{V}, \mathcal{E})$, where $\mathcal{V}$ is the vertex set containing $M$ unique vertices, and $\mathcal{E}$ is the hyperedge set containing $N$ hyperedges. Each hyperedge $\epsilon \in \mathcal{E}$ contains multiple vertices. The hypergraph can be represented by an incidence matrix $\boldsymbol{H}\in \mathbb{R}^{M \times N}$, where $\boldsymbol{h}_{v\epsilon}=1$ if the hyperedge $\epsilon$ contains the vertex $v \in V$, otherwise $\boldsymbol{h}_{v\epsilon}=0$. For each hypergraph, we use the diagonal matrix $\boldsymbol{D}$ to denote the degrees of vertex, where $\boldsymbol{d}_{v}=\sum _{\epsilon =1}^{N}\boldsymbol{w}_{ \epsilon}\boldsymbol{h}_{v\epsilon}$. $\boldsymbol{W} \in \mathbb{R}^{
N \times N}$ is the diagonal matrix of the weight of the hyperedge. Let diagonal matrix $\boldsymbol{B}$ denote the degree of hyperedge, where $\boldsymbol{b}_{\epsilon}=\sum _{v =1}^{M}\boldsymbol{h}_{v\epsilon}$ represents the number of vertices connected by the hyperedge $\epsilon$.

\subsection{Task of Group Recommendation}
The task of the group recommendation is to predict the item ranking list for the given group that has multiple members. We consider $\mathcal{U}=\left \{ u_{1}, u_{2}, ..., u_{|\mathcal{U}|}\right \}$ and $\mathcal{I}=\left \{ i_{1}, i_{2},..., i_{|\mathcal{I}|}\right \}$ to denote the user set and item set, respectively. Let $\mathcal{G}=\left \{ g_{1}, g_{2}, ..., g_{|\mathcal{G}|} \right \}$ be the group set, where $g_{n}\in \mathcal{G}$ is the $n$-th group. There are three observable interaction behaviors: user-item interaction, group-item interaction, and user-group interaction. Let $\boldsymbol{R}\in \mathbb{R}^{|\mathcal{U}|\times |\mathcal{I}|}$ denote the user-item interaction matrix, where $r_{ui}=1$ if the user $u$ consumed item $i$, otherwise $r_{ui}=0$. $\boldsymbol{S}\in \mathbb{R}^{|\mathcal{G}|\times |\mathcal{I}|}$ denotes the group-item interaction matrix, where $s_{gi}=1$ if the group $g$ consumed the item $i$, otherwise $s_{gi}=0$. We embed each user $u$ into the space vector $\boldsymbol{p}_{u} \in \boldsymbol{P}^{|\mathcal{U}|\times d}$ of the dimension $d$. Let $\boldsymbol{q}_{i} \in \boldsymbol{Q}^{|\mathcal{I}|\times d}$ represent the vector representation of item $i$, which is randomly initialized with a $d$-dimensional vector. $\boldsymbol{z}_{g} \in \boldsymbol{Z}^{|\mathcal{G}|\times d}$ denotes the representation of group $g$. We use bold capital letters (e.g., $\boldsymbol{X}$) and bold lowercase letters (e.g., $\boldsymbol{x}$) to represent matrices and vectors, respectively.

\section{The Proposed Method}
In this section, we first introduce the details of the proposed model \textbf{HHGR} (short for "\textbf{H}ierarchical \textbf{H}ypergraph Learning Framework for \textbf{G}roup \textbf{R}ecommendation"). Then, we present its \textbf{s}elf-\textbf{s}upervised variant, $\mathbf{S^{2}}$-\textbf{HHGR}, with the double-scale dropout strategy. Finally, we conduct the complexity analysis on $S^{2}$-HHGR to demonstrate its scalability. The proposed framework is shown in Figure~\ref{f2}, which includes three components: 1) Hierarchical hypergraph, which captures the user interactions within and beyond groups by propagating information from user level to group level; 2) Double-scale self-supervised learning, which contains coarse- and fine-grained node dropout strategies to refine the user and group representations and alleviate the data sparsity problem; and 3) Model optimization, which unifies the objectives of group recommendation and self-supervised learning to enhance both tasks.

\subsection{Hierarchical Hypergraph Convolution}
By organizing users and groups in a hierarchical hypergraph, we can leverage their connectives to help exhibit high-order tuplewise user interactions within and beyond groups, which is of crucial importance to group recommendation. Inspired by \cite{DBLP:conf/aaai/FengYZJG19, DBLP:conf/ijcai/JiangWFCG19}, we devise a new hierarchical hypergraph convolution to perform the embedding propagation mechanism over our hierarchical hypergraph on two levels, user-level and group-level. In what follows, we elaborate on these two ingredients.

\subsubsection{User-level hypergraph}
\textbf{Hypergraph construction.}
Since users are correlated within and beyond groups in group recommendation scenario, it is vital to define appropriate connections among them and exploit the users' mutual interactions. The conventional graph structure can only support pairwise relations between users, which is not fit for this case. Thus, we propose to model such user interactions with a hypergraph, which is shown in the bottom-left corner of Figure~\ref{f2}. In the user-level hypergraph $G_{ul}$, multiple users can be connected through one hyperedge (or group). 

\textbf{User representation learning.}
We aim to exploit the tuplewise user interactions for learning their influences, in which the correlated users should be affected by the intra-group users. To achieve that, we introduce the hypergraph convolution operation to capture user interactions among their neighbors and learn each user's dynamic representation. For the specific user node $v_{u}$, the hypergraph convolutional network first learns the representations of all its connected hyperedges $\epsilon_{j}$, which is from the gathered node features. Then the node $v_{u}$'s representation would integrate the related hyperedge feature information. Simple graphs use the adjacency matrix to represent relationships between two connected nodes, whereas hypergraphs introduce the incidence matrix $\bm{H}_{ul}$ to describe the relationship between nodes and hyperedges, which is defined in Section 2.1. Referring to the spectral hypergraph convolution proposed in HGNN \cite{DBLP:conf/aaai/FengYZJG19, wu2019simplifying}, the hypergraph convolution is defined as follows:
\begin{equation}
\label{eq:eq1}
\bm{P}^{(l+1)}=\sigma (\bm{D}_{ul}^{-1}\bm{H}_{ul}\bm{W}_{ul}\bm{B}_{ul}^{-1}\bm{H}_{ul}^{T}\bm{P}^{(l)}\bm{\Theta}^{(l)}),
\end{equation}
where $\bm{D}_{ul}$ denotes the vertex degree matrix of the user-level hypergraph, and $\bm{B}_{ul}$ denotes the hyperedge degree matrix. $\bm{W}_{ul}$ is regarded as the weight matrix of the hyperedges. In this paper, we initialize the weight matrix $\bm{W}_{ul}$ with the identity matrix yielding equal weights for all hyperedges. $\bm{D}_{ul}$ and $\bm{B}_{ul}$ play a role of the normalization in hypergraph convolutional operator. $\bm{\Theta}$ is the parameter matrix between two convolutional layers. $\bm{P}^{(l)}$ is the users' embeddings in the $l$-th hypergraph convolutional network, where $\bm{P}^{(0)}=\bm{\tilde{P}}$. $\bm{\tilde{P}}$ is the initial vectors of all users $u$ with $d$-dimension. Since the nonlinear activation is found to be redundant in recent research \cite{wu2019simplifying}, we remove this part in Eq. (1), which is shown as follows:
\begin{equation}
\label{eq:eq2}
\bm{P}^{(l+1)}= \bm{D}_{ul}^{-1}\bm{H}_{ul}\bm{B}_{ul}^{-1}\bm{H}_{ul}^{T}\bm{P}^{(l)}\bm{\Theta}^{(l)}.
\end{equation}

\textbf{Group representation learning.}
Our target is to obtain the group embeddings to estimate their preferences on items. Since group members have different importance, we perform a weighted sum on the representations of user members to generate the attentive group representation $\bm{z}_{g_{m}} \in \tilde{\bm{Z}}$. The weights can reflect the users' contribution to the group's decision-making.
\begin{equation}
\label{eq:eq3}
\tilde{\bm{z}_{g_{m}}}=\sum_{u \in g_{m}} \alpha_{u} \bm{p}_{u},
\end{equation}

\begin{equation}
\label{eq:eq4}
\alpha_{u}=\frac{\exp \left(\bm{p}_{u} \bm{W}_{\mathrm{agg}} \bm{x}^{T} \right)}{\sum_{{j} \in g_{m}} \exp \left(\bm{p}_{j} \bm{W}_{\mathrm{agg}} \bm{x}^{T} \right)},
\end{equation}
where $\alpha_{u}$ is the weight of a user $u$ in the group decision, $\bm{p}_{j}$ denotes other members vectors in the same group $g_{m}$, and $\bm{W}_{agg}\in\mathbb{R}^{d\times d}$ and $\bm{x}\in\mathbb{R}^{d}$ are parameters used to compute the weights.

\subsubsection{Group-level hypergraph}
\textbf{Hypergraph construction.}
A user often belongs to multiple groups, which means she is connected with other users of different groups. However, not all user interactions affect user preferences. To select informative user interactions and optimize inter-group interactions among users, we adopt the triadic motif relation as the motif-induced hyperedge to construct the group-level hypergraph. Since the triadic motif can increase homophily cohesion in the social community \cite{DBLP:journals/pvldb/LeeKS20, DBLP:conf/socialcom/FriggeriCF11}, its interaction is a stronger bond, which is highly beneficial to modeling user interactions. Specifically, we first convert the user-level hypergraph to a projected graph, where group hyperedges act as nodes. In the projected graph, if two group hyperedges have common user members, they will be connected, ensuring the relevance of preferences between groups. Then, we adopt the triadic motif to select the most relevant groups in the group-level hypergraph. If any three groups conform to the definition of the motif, we will define these three groups to be divided into one motif-induced hyperedge. The group-level hypergraph is represented as $G_{gl}$.

\textbf{Group representation learning.}
Intuitively, the motif can be regarded as a closed path connecting three different nodes. Referring to  the motif-based PageRank model proposed in \cite{ DBLP:conf/aaai/ZhaoXSLCG18}, when we do not consider the self-connection, motif-based adjacency matrix can be calculated as Eq. (5), which can be represented as $\bm{H}_{gl}\bm{H}_{gl}^{T}$. 
\begin{equation}
\label{eq:eq5}
\bm{H}_{gl}\bm{H}_{gl}^{T} = \left ( \bm{C}\bm{C}\right )\odot\bm{C},
\end{equation}
where $\bm{H}_{gl} \in \mathbb{R}^{\left | \mathcal{G} \right |\times \left | \mathcal{G} \right |}$ denotes the motif incidence matrix of the group-level hypergraph. $\bm{C} \in \mathbb{R}^{\left | \mathcal{G} \right |\times \left | \mathcal{G} \right |}$ is the symmetric adjacency matrix of the projected graph, where $c_{ij}=1$ if there is a connection between group $i$ and group $j$, otherwise $c_{ij}=0$. $\bm{C}\bm{C}$ denotes the paths connecting three vertices, and $\odot\ \bm{C}$ transforms the paths into  closed triangles.

Similar to the hypergraph convolution proposed in the user-level hypergraph, the transformed hypergraph convolutions in the group-level hypergraph can be defined as Eq. (6). Following the LightGCN and MHCN \cite{DBLP:conf/sigir/0001DWLZ020, DBLP:conf/www/YuYLWH021}, since the self-connection has little effect on performance, the Eq. (6) is equivalent to Eq. (2).
\begin{equation}
\label{eq:eq6}
\bm{Z}^{(l+1)}=\bm{D}_{gl}^{-1} \bm{H}_{gl}\bm{H}_{gl}^{T} \bm{Z}^{(l)}\bm{\Psi }^{(l)},
\end{equation} 
where $\bm{D}_{gl} \in \mathbb{R}^{\left | \mathcal{G} \right |\times \left | \mathcal{G} \right |}$ is the degree matrix of the hypergraph motif $\bm{H}_{gl}\bm{H}_{gl}^{T}$. $\bm{H}_{gl}\bm{H}_{gl}^{T}$ can be regarded as the motif-induced adjacency matrix.  $\bm{\Psi}^{(l)}$ is the parameter matrix in the $l$-th layer. $\bm{Z}^{(l)}$ is the group representation in the $l$-th layer of the hypergraph convolutional network, where $\bm{Z}^{(0)}=\tilde{\bm{Z}}$. The $\tilde{\bm{Z}}$ learns from the attention-based group preference aggregator. 

\subsubsection{Loss function}
Here, we adopt the pairwise learning task loss function to optimize user and item representations in the user-level hypergraph, which is designed as follows:
\begin{equation}
\label{eq:eq7}
\begin{aligned}
& \mathcal{L}_{UI}=\sum_{(u, i, j) \in O}\left(\hat{r}_{ui}-\hat{r}_{uj}-1\right)^{2}, \\
& \hat{r}_{ui} = \sigma\left( \bm{p}_{u} \bm{q}_{i}^{T} \right),
\end{aligned}
\end{equation}
where $\bm{p}_{u}$ represents the embedding vector for user $u$. $\bm{q}_{i}$ represents the embedding vector for item $i$. $O$ represents the training set, in which each one includes the user $u$, interacted item $i$, and unobserved items $j$. Our target is to make the margin between the positive samples $(u, i)$ and negative samples $(u, j)$ is as close to 1 as possible.

For the group-level hypergraph, its loss function can be achieved as follows:
\begin{equation}
\label{eq:eq8}
\begin{aligned}
 & \mathcal{L}_{GI} = \sum_{(g, i, j) \in O^{'}}\left(\hat{s}_{gi}-\hat{s}_{gj}-1\right)^{2}, \\
&  \hat{s}_{gi}= \sigma\left( \bm{z}_{g} \bm{q}_{i}^{T} \right),
\end{aligned}
\end{equation}
where $\bm{z}_{g}$ is the group representation learned from the attention-based preference aggregation strategy. Similar to the user-level loss, $O^{'}$ represents the training set, which includes the group $g$, interacted item $i$, and unobserved items $j$. To enhance the learning process of these two recommendation methods, we apply a joint training strategy, whose loss function consists of two parts: user preference loss and group preference loss. It is given by
\begin{equation}
\label{eq:eq9}
\mathcal{L}=\mathcal{L}_{UI} + \mathcal{L}_{GI}.
\end{equation}

\subsection{Boosting Group Recommendation with Double-Scale Self-Supervision}
Despite the remarkable capability of hypergraph modeling, as groups may occasionally form, only minimal group-item interactions can be observed (i.e., the data sparsity issue), which may lead to sub-optimal recommendation performance \cite{DBLP:conf/icde/YinW0LYZ19, DBLP:journals/corr/abs-2103-13506, DBLP:conf/sigir/HeCZ20}. Several models utilize the user member preferences to synthesize group preferences and alleviate the sparsity issue. However, they overlook the fact that the user-item interaction is also sparse. To address this problem and refine the representations of users and groups, we propose a double-scale node dropout strategy on the hierarchical hypergraph (including coarse and fine granularity) to augment the raw data and create two types of self-supervision signals for contrastive learning to boost the performance of HHGR. A self-supervised variant of HHGR, $S^{2}$-HHGR, is presented as follows. It should be noted that, for $S^{2}$-HHGR, we upgrade the design in Section 3.1.1 and employ two new hypergraph convolutional networks to learn user representations, which can be seamlessly coupled with the double-scale self-supervised strategy. 

\textbf{Coarse-grained node dropping strategy.} We drop a certain portion of users at a coarse-grained granularity, where the magnitude of missing nodes is a hyperparameter. Unlike node dropping in a simple graph accompanied by deleting connected edges, the hyperedge may not be deleted when discarding some nodes in the hypergraph. Besides, when a deleted node belongs to multiple group hyperedges, the node would be deleted in all certain hyperedges. We define the coarse-grained node dropping function $f_{coarse}$, as follows:
\begin{equation}
\label{eq:eq10}
\bm{h}_{c}=f_{coarse}(\bm{a}_{c}, \bm{h}_{ul})=\bm{a}_{c} \odot \bm{h}_{ul},
\end{equation}
where $\bm{h}_{c}$ represents the column vector of the coarse-grained hypergraph incidence matrix $\bm{H}_{c}$, and $\bm{a}_{c}\in\{0,1\}^{|\mathcal{U}|}$ is a mask vector with its entries being 0 at a given probability, controlling the dropout magnitude of the nodes in $\bm{H}_{ul}$. $\odot$ represents the element-wise product, which means the mask vector would multiply with each column in the user-level hypergraph incidence matrix $\bm{H}_{ul}$. After the coarse-grained node dropping, we can obtain a perturbed user-level hypergraph. We encode it through a new hypergraph convolutional network $g_{c}(\cdot)$ to get the coarse-grained user representation $\bm{P}'$, which can be as one data augmentation of the raw user representations.

\begin{equation}
\label{eq:eq11}
\bm{P}'^{(l+1)}=g_{c}(\bm{P}'^{(l)})=\bm{D}_{c}^{-1}\bm{H}_{c}\bm{B}_{c}^{-1}\bm{H}_{c}^{T}\bm{P}'^{(l)}\bm{\Gamma }^{(l)},
\end{equation}
where $\bm{D_{c}}$ and $\bm{B_{c}}$ are the vertex degree and hyperedge degree diagonal matrices of the coarse-grained hypergraph. $\bm{\Gamma}$ denotes the parameter matrix in the coarse-grained hypergraph convolutional network. $\bm{{P}'}^{(l)}$ is the user representation in the $l$-th layer of the coarse-grained hypergraph neural network, where $\bm{{P}'}^{(0)} = \bm{\tilde{P}}$. $ \bm{\tilde{P}}$ is the randomly initialized d-dimensional representation matrix.

\textbf{Fine-grained node dropping strategy.} We perturb the hypergraph construction by dropping a certain number of users in a particular group hyperedge. Analogously, we define a fine-grained node dropping function $f_{fine}$, which is represented as follows:
\begin{equation}
\label{eq:eq12}
\bm{h}_{f}=f_{fine}(\bm{a}_{f}, \bm{h}_{ul})=\bm{a}_{f} \odot \bm{h}_{ul},
\end{equation}
where $\bm{h}_{f}$ represents the column vector of the fine-grained hypergraph incidence matrix $\bm{H}_{f}$, and $\bm{a}_{f}\in\{0,1\}^{|\mathcal{U}|}$ is the fine-grained mask vector, which will drop nodes with a fixed probability in only one hyperedge. Different from the coarse-grained strategy, each time $\bm{a}_{f}$ is multiplied by $\bm{h}_{f}$, it is reassigned. Next, we encode $\bm{P}''^{(l)}$ through $g_{f}(\cdot)$ to obtain a new user representation:
\begin{equation}
\label{eq:eq13}
\bm{P}''^{(l+1)} = g_{f}(\bm{P}''^{(l)}) = \bm{D}_{f}^{-1}\bm{H}_{f}\bm{B}_{f}^{-1}\bm{H}_{f}^{T}\bm{P}''^{(l)}\bm{\Phi }^{(l)},
\end{equation}
where $\bm{D_{f}}$ and $\bm{B_{f}}$ are the vertex and hyperedge degree diagonal matrices of the fine-grained hypergraph convolutional network. $\bm{H_{f}}$ is the incidence matrix. $\bm{\Phi }$ represents the parameter matrix. $\bm{{P}''}^{(l)}$ is the user representation in the $l$-th layer of the fine-grained hypergraph neural network, where $\bm{{P}''}^{(0)} = \bm{\tilde{P}}$. To avoid the loss of some essential information, we add two representations at the two granulates to get the final user representation: $\bm{P}=\bm{{P}'}+\bm{{P}''}$.

\textbf{Contrastive learning.} Having established different granularity augmented views of nodes, we treat the granularity of the same node as the positive pairs and any different granularity of nodes as the negative pairs. We hope that the distribution of the user representation vectors from two pretext tasks can be as close as possible. Hence, we design a discriminator function $\mathit{f}_{\mathcal{D}}(\cdot)$ to learn a score between two input vectors and assign higher scores to positive pairs compared with negative samples. We adopt the cross-entropy loss as contrastive loss function to enforce maximizing the agreement between positive pairs, which is defined as follows:
\begin{equation}
\label{eq:eq14}
\begin{aligned}
\mathcal{L}_{UU}=&-\sum_{i \in U}\left[\log \left(\mathit{f}_{\mathcal{D}}\left(\bm{{p}'}_{i}, \bm{{p}''}_{i}\right)\right) + \sum_{j=1}^{n}\left[\log \left(1-\mathit{f}_{\mathcal{D}}\left(\bm{{p}'}_{j}, \bm{{p}''}_{i}\right)\right) \right]\right],
\end{aligned}
\end{equation}

\begin{equation}
\label{eq:eq15}
\begin{aligned}
\mathit{f}_{\mathcal{D}} \left(\bm{{p}'}_{i}, \bm{{p}''}_{i} \right)= \sigma\left(\bm{{p}'}_{i} \bm{W}_{\mathcal{D}} \bm{{p}''}^{T}_{i} \right),
\end{aligned}
\end{equation}
where $\bm{{p}'}_{j}$ is the representation of user $j$ in the coarse-grained hypergraph; $n$ is the number of negative samples randomly selected from the same batch. $\bm{W}_{\mathcal{D}}$ is weight parameters of discriminator.

\subsection{Model Optimization}
To optimize the $S^{2}$-HHGR's parameters, we unify the loss functions of the task of group recommendation and self-supervised learning. We first update the loss function of self-supervised learning, $\mathcal{L}_{UU}$, as the pre-training strategy to optimize the user representations. Then we optimize the supervised loss function of $\mathcal{L}_{UI}$ and $\mathcal{L}_{GI}$ to obtain the representations of users and items. Meanwhile, the model will be fine-tuned and updated during the process of self-supervised learning. The learned user embedding would be used to generate group representations, which are updated in the objective function $\mathcal{L}_{GI}$. Specifically, we train our model using the Adam algorithm. The overall objective is defined as follows:
\begin{equation}
\label{eq:eq16}
\mathcal{L}=\beta \mathcal{L}_{UU}+\mathcal{L}_{UI}+\mathcal{L}_{GI},
\end{equation}
where $\beta$ is the hyper-parameter to balance the task of self-supervised learning and supervised learning task for group recommendation.

\subsection{Complexity Analysis}
In this section, we discuss the complexity of $S^{2}$-HHGR from model size and time complexity.

\textbf{Model size.}
The parameters of our model consist of three parts: 1) hierarchical hypergraph convolutional network, 2) attention-based group preference aggregator, and 3) discriminator network. The coarse-grained and fine-grained user-level hypergraph convolutional network have parameters of size $2[l \times d \times d ]$. The parameters size of the group-level hypergraph convolutional network is $l \times d \times d $. The parameter size of the trainable weighted matrix in the preference aggregator is $3 \times d \times d$. The discriminator measures the similarity between user representations of different granularities, the weight matrix in $\mathit{f}_{\mathcal{D}}(\cdot)$ has a trainable weighted matrix $\bm{W}_{D}$ with the shape ${d \times d}$. 

\textbf{Time complexity.}
The computational cost mainly comes from two parts: the hierarchical hypergraph convolution and the double-scale self-supervised learning strategy. The time complexity of hierarchical hypergraph is mainly from the information propagation consumption, through $l$, is less than $O(l \times |\bm{H}|\times d )$, where $|\bm{H}|$ denotes the number of nonzero elements in the incidence matrix $\bm{H}$. Since our model contains a user-level hypergraph convolutional network and a group-level hypergraph convolutional network, the time complexity of the hierarchical hypergraph convolutional neural network is about $O(l^{2} \times |\bm{H}|^{2} \times d^{2} )$. As for the self-supervised learning strategy, the cost derives from the hypergraph generation of different perspectives and contrastive learning. The time cost of hypergraph generation is less than $O(2 \times M\times M)$. The time cost of contrastive learning is less than $O(|\mathcal{U}|\times n)$, where $n$ is the number of negative samples. 

\begin{table}[ht]
\caption{The statistics of datasets.}
\small
\label{tab:dataset}
\begin{tabular}{llllll}
\hline
\multicolumn{1}{c}{\textbf{Dataset}} & \multicolumn{1}{c}{\textbf{\#User}} & \multicolumn{1}{c}{\textbf{\#Item}} & \multicolumn{1}{c}{\textbf{\#Group}} & \multicolumn{1}{c}{\textbf{\begin{tabular}[c]{@{}c@{}}\#U-I \\ Feedback\end{tabular}}} & \multicolumn{1}{c}{\textbf{\begin{tabular}[c]{@{}c@{}}\#G-I \\ Feedback\end{tabular}}} \\ \hline
\textbf{Douban} & 2,964 & 39,694 & 2,630 & 823,851 & 463,040\\ 
\textbf{Weeplaces} & 8,643 & 25,081 & 22,733 & 1,358,458 & 180,229 \\
\textbf{GAMRa2011} & 602 & 7,710 & 290 & 116,344 & 145,068\\
\hline
\end{tabular}
\end{table}

\begin{table*}[ht]
\caption{The general recommendation performance comparison on three datasets.}
\label{tab:performance}
\begin{tabular}{cllllllllllll}
\hline
\textbf{Dataset} & \multicolumn{4}{c}{\textbf{Weeplaces}} & \multicolumn{4}{c}{\textbf{CAMRa2011}} & \multicolumn{4}{c}{\textbf{Douban}} \\
\textbf{Metric} & \textbf{N@20} & \textbf{N@50} & \textbf{R@20} & \textbf{R@50} & \textbf{N@20} & \textbf{N@50} & \textbf{R@20} & \textbf{R@50} & \textbf{N@20} & \textbf{N@50} & \textbf{R@20} & \textbf{R@50} \\ \hline
\multicolumn{13}{c}{\textbf{Baseline recommender}} \\ \hline
\textbf{Popular} & 0.063 & 0.074 & 0.126 & 0.176 & 0.099 & 0.122 & 0.172 & 0.226 & 0.003 & 0.005 & 0.009 & 0.018 \\ 
\textbf{NeuMF} & 0.193 & 0.244 & 0.271 & 0.295 & 0.305 & 0.367 & 0.393 & 0.450 & 0.124 & 0.167 & 0.248 & 0.316 \\ \hline
\multicolumn{13}{c}{\textbf{Attention-based group recommender}} \\ \hline
\textbf{AGREE} & 0.224 & 0.267 & 0.354 & 0.671 & 0.307 & 0.418 & 0.529 & 0.688 & 0.201 & 0.224 & 0.297 & 0.488 \\
\textbf{MoSAN} & 0.287 & 0.334 & 0.548 & 0.738 & 0.423 & 0.466 & 0.572 & 0.801 & 0.163 & 0.209 & 0.384 & 0.459 \\
\textbf{SIGR} & 0.357 & 0.391 & 0.524 & 0.756 & 0.499 & 0.524 & 0.585 & 0.825 & 0.217 & 0.235 & 0.436 & 0.560 \\
\textbf{GroupIM} & 0.431 & 0.456 & 0.575 & 0.773 & 0.637 & 0.659 & 0.753 & 0.874 & 0.257 & 0.284 & 0.523 & 0.696 \\
\textbf{HHGR} & 0.379 & 0.398 & 0.554 & 0.764 & 0.517 & 0.532 & 0.703 & 0.830 & 0.254 & 0.267 & 0.507 & 0.677 \\
$\mathbf{S^{2}}$-HHGR & \textbf{0.456} & \textbf{0.478} & \textbf{0.592} & \textbf{0.797} & \textbf{0.645} & \textbf{0.671} & \textbf{0.779} & \textbf{0.883} & \textbf{0.279} & \textbf{0.294} & \textbf{0.561} & \textbf{0.741} \\ \hline
\end{tabular}
\end{table*}

\section{Experience and Results}
In this section, we conduct extensive experiments to justify our model's superiority and reveal the reasons for its effectiveness. Specifically, we will answer the following research questions to unfold the experiments.

\textbf{RQ1:} Compared with the state-of-the-art group recommendation models, how does our model perform?

\textbf{RQ2:} What are the benefits of each component (i.e., the hierarchical hypergraph and the self-supervised learning) in our model? 

\textbf{RQ3:} How do the hyper-parameters influence the effectiveness of the $S^{2}$-HHGR?

\subsection{Experimental Settings}
\subsubsection{Datasets.} We conduct experiments on three public datasets: Weeplaces \cite{DBLP:conf/sigir/SankarWWZYS20}, CAMRa2011 \cite{DBLP:conf/cikm/LiuWSM14}, and Douban \cite{DBLP:conf/icde/YinW0LYZ19}. Weeplaces dataset includes the users' check-in history in a location-based social network. We follow GroupIM \cite{DBLP:conf/sigir/SankarWWZYS20} to construct group interactions by using the user check-in records and their social network. As for the CAMRa2011, it is a movie rating dataset containing individual users and households records. We follow the idea of AGREE \cite{DBLP:conf/sigir/Cao0MAYH18} and convert the explicit rating to implicit preference, where the rating records are regarded as 1. Douban dataset is from the Douban platform, which consists of a variety of group social activities. As the Douban dataset does not contain explicit group information, we follow the idea of SIGR \cite{DBLP:conf/icde/YinW0LYZ19} to extract implicit group data. Specifically, we regard users' friends who have participated in the same activity as the group members. The statistical information of the three datasets is shown in Table~\ref{tab:dataset}. We randomly split the set of all groups into training $(70\%)$, validation $(10\%)$, and test sets $(20\%)$.

\subsubsection{Baselines.} To answer the \textbf{RQ1}, we compare the proposed model with the following models:
\begin{itemize}[leftmargin=*]
\item \textbf{Popularity}. This method ranks items according to their popularity.
\item \textbf{NeuMF} \cite{DBLP:conf/www/HeLZNHC17}. NeuMF is a neural network-based collaborative filtering model to benchmark the recommendation performance. We treat all groups as virtual users and utilize group-item interactions to generate the group preferences. 
\item \textbf{AGREE} \cite{DBLP:conf/sigir/Cao0MAYH18}. This method utilizes attentional preference aggregation to compute group member weights and adopts neural collaborative filtering to learn the group-item interaction.
\item \textbf{MoSAN} \cite{DBLP:conf/sigir/TranPTLCL19}. MoSAN is a neural group recommender that employs a collection of sub-attentional networks to learn each user's preference and model member interactions. 
\item \textbf{SIGR} \cite{DBLP:conf/icde/YinW0LYZ19}. This is a state-of-the-art group recommendation model, which introduces a latent variable and the attention mechanism to learn users' local and global social influence. It also utilizes the bipartite graph embedding model to alleviate the data sparsity problem.
\item \textbf{GroupIM} \cite{DBLP:conf/sigir/SankarWWZYS20}. This model aggregates the users' preferences as the group preferences via the attention mechanism. It maximizes the mutual information between the user representations and its belonged group representations to alleviate the data sparsity problem.
\end{itemize}
\textbf{HHGR} is the vanilla version of our proposed model, and $\mathbf{S^{2}-HHGR}$ represents the self-supervised version.

\subsubsection{Evaluation metrics.} To measure the performance of all methods, we employ the widely adopted metrics $NDCG@K$ and $Recall@K$ with $k=\left \{ 20,50 \right \} $. $NDCG@K$ evaluates the ranking of true items in the recommendation list. $Recall@K$ is the fraction of relevant items that have been retrieved in the Top-K relevant items.

\subsubsection{Parameter settings.} For the general settings, the embedding size is 64, the batch size for the mini-batch is 512, and the number of negative samples is 10. During the training process of the double-scale self-supervised learning, the initial learning rate is $5e-4$. For the group-level hypergraph training, the learning rate is $1e-4$. The user-level and group-level hypergraph neural network structure is two-layer and one-layer, respectively. For the baseline models, we refer to their best parameter setups reported in the original papers and directly report their results if we use the same datasets and evaluation settings.

\subsection{Recommendation Performance Comparison (\textbf{RQ1})}
\subsubsection{Overall performance comparison.}
In this part, we validate the superiority of HHGR and $S^{2}$-HHGR on three datasets. Table~\ref{tab:performance} shows the experimental results of the proposed models' performance compared with the baselines. We highlight the best results of all models in boldface. According to the results, we note the following key observations:

1) Among these methods, the attention-based group models outperform baseline recommenders (i.e., Popular and NeuMF) on most datasets due to their ability to dynamically model the user interactions within groups and learn various weights of different members in the group. 2) In the models based on attention mechanism, HHGR is better than most models (including AGREE, MoSAN, and SIGR). We believe that the performance improvement of HHGR verifies the effectiveness of hypergraph and hypergraph neural network modules to exploit high-order user interactions. 3) SIGR performs better than AGREE and MoSAN due to considering the form of bipartite graph to represent user-item interactions, especially on the CAMRa2011 dataset. Besides, MoSAN also achieves better results because of its expressive power of preference aggregators to capture different personal weights in group-item interactions. 4) On the other hand, although HHGR is slightly inferior to GroupIM, the enhanced model $S^{2}$-HHGR beats the most advanced group recommendation models on three datasets, which verifies the effectiveness of self-supervised learning strategies.

\begin{figure}[ht]
\includegraphics[width=\linewidth]{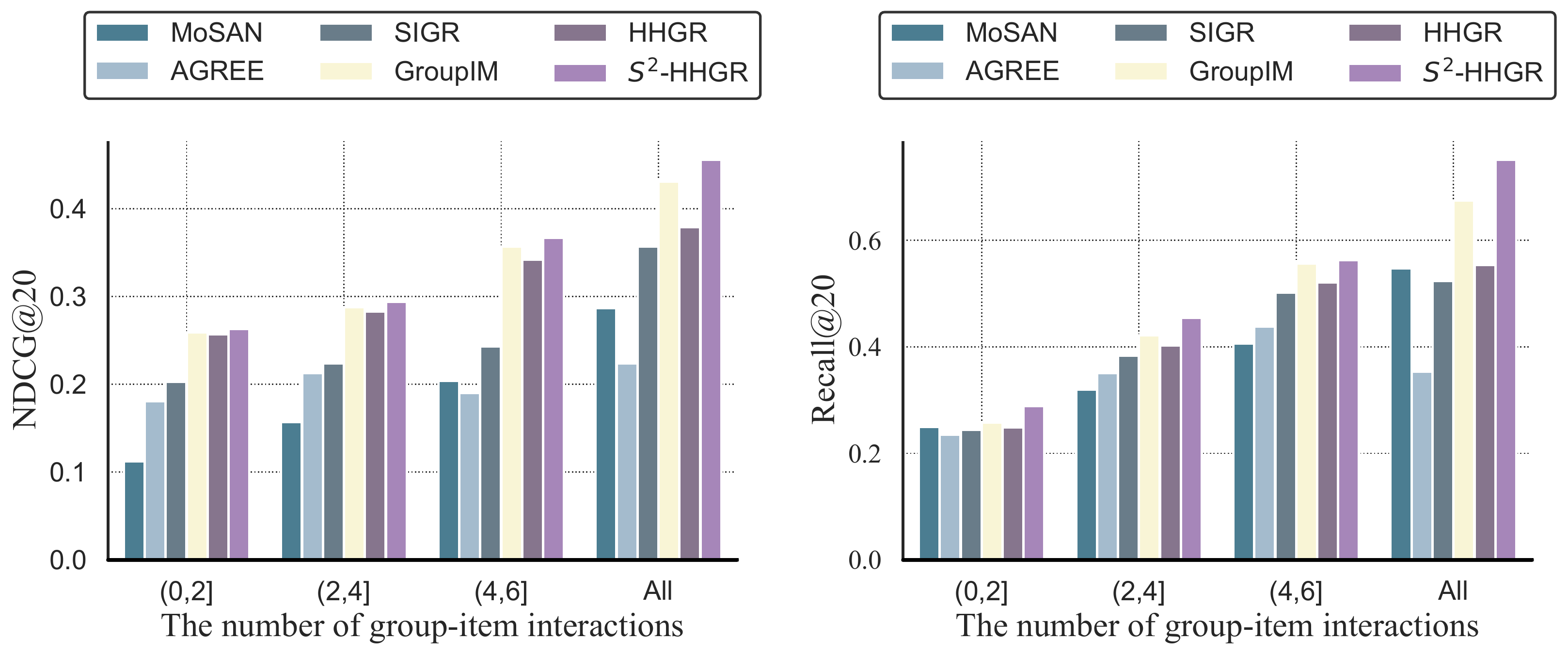}
\caption{Performance comparison of attention-based group recommendation models on sparsity datasets.}
\label{f4}
\end{figure}

\subsubsection{Performance on sparsity datasets.}
We adopt the self-supervised learning strategy to alleviate the data sparsity problem. To validate the effectiveness of the proposed double-scale node dropout strategy, we study experiments on sparse datasets. We split the dataset into four groups based on the number of group-item interactions. We find that the performance of all models varies with the number of group interactions. Due to the limited space, we only show the performance of attention-based models on the Weeplaces dataset. The results are shown in Figure~\ref{f4}. 

As shown in Figure~\ref{f4}, NDCG and Recall generally increase as the number of interacting groups of an item increases, since more interactions will provide more information for learning the group preference representation. Without the auxiliary information, models that adopt self-supervised learning (GroupIM and $S^{2}$-HHGR) are competitive to the models without self-supervised learning. Besides, we can observe that $S^{2}$-HHGR shows the best performance compared with other models. We argue that the effectiveness of $S^{2}$-HHGR can be attributed to the double-scale node dropout strategy in hypergraph neural networks, which can enhance the representations of group preferences. 

\subsection{Ablation Study (\textbf{RQ2})} 
\subsubsection{Investigation of the hierarchical hypergraph}
To investigate the effectiveness of the hierarchical hypergraph, we conduct experiments on Weeplaces and CAMRa2011 for two HHGR variants, each of which has one of the level removed. \textbf{HHGR-wg} and \textbf{HHGR-wu} to denote the ablated model without group level or user level. HHGR-wg only considers the user-level hypergraph, which means the group preference representations generated from the preference aggregator are the final group preference. HHGR-wu adopts the group-item interaction information to initialize the group preference and only considers the group level hypergraph to update the group preference. The experimental results are demonstrated in Table~\ref{tab:hypergraph}. From the results, we can observe that the performance of HHGR-wu falls to the performance of HHGR-wg and HHGR on two datasets. Compared with the Weeplace dataset, the performance of the HHGR and its variants on the CAMRa2011 is better. The possible reason might be that the CAMRa2011 dataset is denser, which is beneficial to selecting informative inter- and intra-group user interactions.

\begin{table}[ht]
\caption{Comparison between HHGR and its variants.}
\label{tab:hypergraph}
\begin{tabular}{l|llll}
\hline
\textbf{Method} & \multicolumn{2}{c}{\textbf{Weeplaces}} & \multicolumn{2}{c}{\textbf{CAMRa2011}} \\
\textbf{Metric} & \textbf{N@50}      & \textbf{R@50}     & \textbf{N@50}      & \textbf{R@50}     \\ \hline
HHGR-wu & 0.288  & 0.683  & 0.495    & 0.797             \\
HHGR-wg  & 0.378   & 0.751  & 0.511 & 0.815             \\
HHGR & 0.398    & 0.764  & 0.532    & 0.830             \\ \hline
\end{tabular}
\end{table}

\subsubsection{Investigation of self-supervised learning}
\textbf{Different types of self-supervised learning.} To investigate the feasibility and efficiency of self-supervised learning, we conduct an ablation study to investigate each component's contributions. We propose two variants of $S^{2}$-HHGR: \textbf{HHGR-F} and \textbf{HHGR-C}. \textbf{HHGR-F} means that only fine-grained node dropping is considered. \textbf{HHGR-C} means only coarse-grained node dropping is used. For the above two variant models, we maximize the mutual information between the initial node representations and the dropping ones. Considering the limited space of the paper, we only compare these models with $S^{2}$-HHGR on Weeplaces and CAMRa2011.

\begin{figure}[ht]
\includegraphics[width=\linewidth]{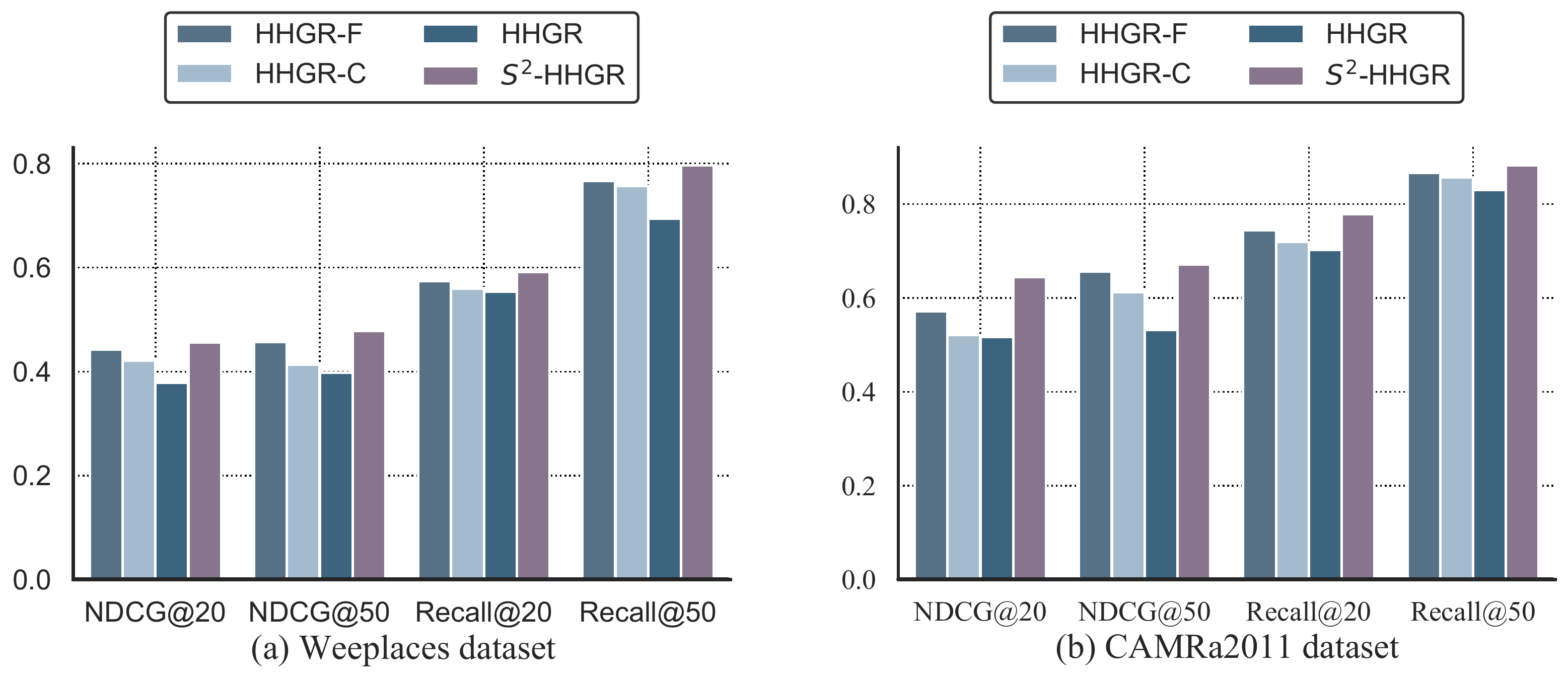}
\caption{The influence of different self-supervised learning strategies.}
\label{f5}
\end{figure}

Figure~\ref{f5} shows that $S^{2}$-HHGR achieves the best performance over independent granularity models, where self-supervised learning strategies contribute to the recommendation performance. When we only use fine-grained node dropping, the HHGR-F performs slightly worse than $S^{2}$-HHGR but better than HHGR-C, which implies the fine-grained node dropping contributes more to the node representation learning. We argue that the fine-grained task would generate more hard samples, strengthening model training. When we only use the coarse-grained node dropout strategy, the group recommendation performance may be slightly better than HHGR. Besides, without the self-supervised learning strategy would result in a performance decline in most cases on two datasets. The performance of HHGR is worse than HHGR-F and HHGR-C in most cases, indicating that the self-supervised strategy is effective. Overall, the experimental results show that self-supervised learning is useful.

\begin{figure}[ht]
\includegraphics[width=\linewidth]{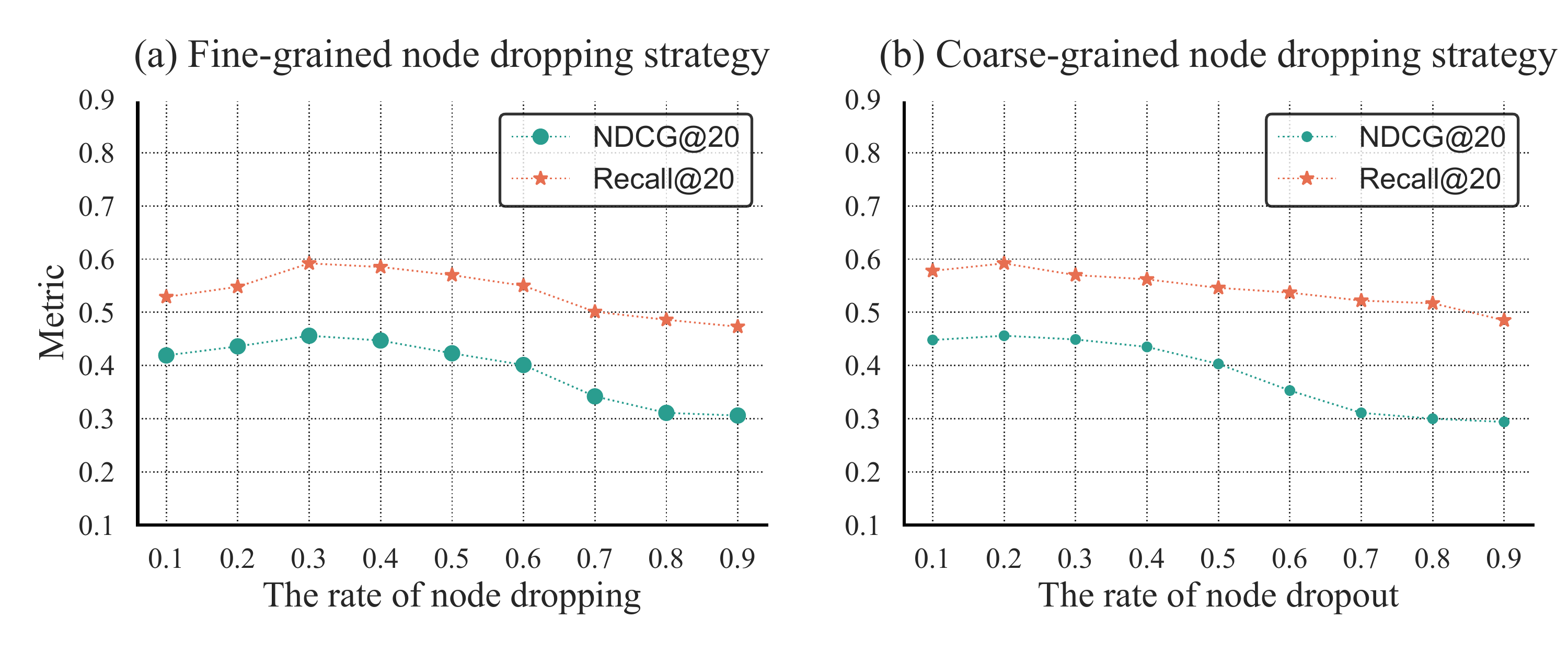}
\caption{The performance of different extent of two pretext tasks.}
\label{f6}
\end{figure}

\textbf{Different extent of self-supervised learning.} We note that the use of different granularities of node dropping benefits the node representation. We further analyze the extent of coarse and fine granularity node dropout. Specifically, we set different dropping rates from 0.1 to 0.9 in two granularity tasks. When we change the rate of one granularity task, we will fix another granularity's rate. As presented in Figure~\ref{f6}, with the increase of dropping rate, there is a significant increase in performance. When the rate reaches the peak, approximately 0.3 and 0.2 on fine and coarse granularity dropping, respectively, it decreases. It suggests that node dropping can improve the HHGR performance with increasing dropping strength, but too many deletions will worsen the data sparsity problem. 

\subsection{Parameter Sensitivity Analysis (\textbf{RQ3})}
This part focuses on pointing out which hyper-parameters affect our model. The analyzed hyper-parameters include the learning rate, the depth of hypergraph convolutional layers, the number of batch sizes, and the number of negative samples. Due to limited space, we only show the experimental results on the Weeplaces dataset. Specifically, we search the proper value in a small interval and set the learning rate as $\left \{1e-5, 5e-5, 1e-4, 5e-4, 1e-3\right \}$, and report the performance of $S^{2}$-HHGR with five values $\left \{ 1,2,3,4,5 \right \}$ of the hypergraph convolutional layer. Besides, the batch size changes from 16 to 512 at intervals of increasing powers of 2, and the number of negative samples is set from 5 to 30 at intervals of 5.

\begin{figure}[ht]
\includegraphics[width=\linewidth]{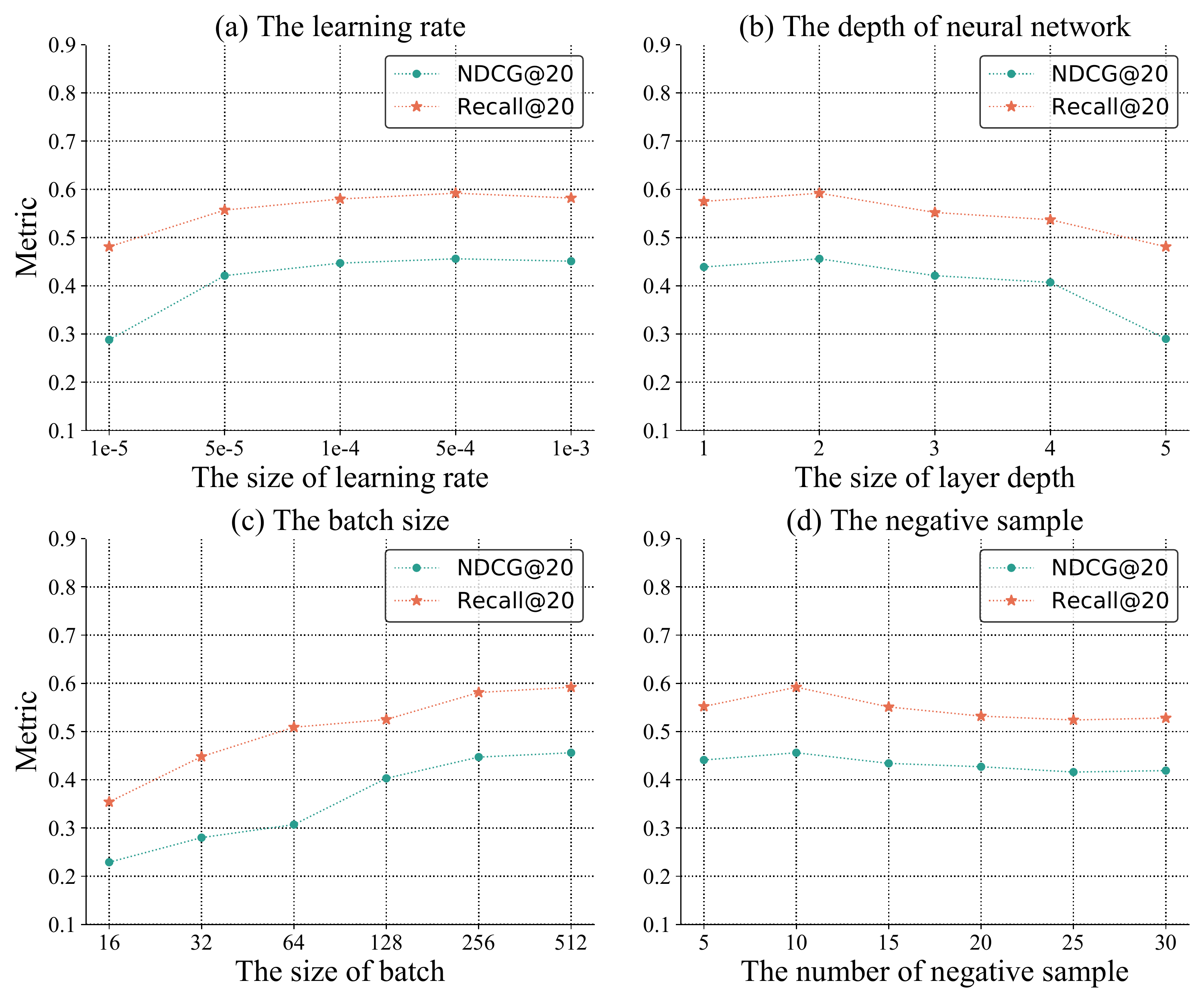}
\caption{The influence of the model parameters.}
\label{f7}
\end{figure}

As shown in Figure~\ref{f7} (a), we observe $5e-4$ is sufficient for the learning rate. With the increase of the rate, the performance shows a trend of rising first and then falling, reaching the peak at $5e-4$. According to Figure~\ref{f7} (b), when the depth of hypergraph convolutional layers is 2, the model reaches the performance peak. As the depth of convolutional layers increases, the performance of $S^{2}$-HHGR steadily declines. A possible reason is that a multi-layer hypergraph convolutional network would lead to the over-smoothing problem. HHGR model also encounters this problem, but it tends to over-smoothing at lower layers. We believe that a self-supervised learning strategy can alleviate the data sparsity problem because it would delete some redundant information to enhance the model performance. Besides, the parameter of batch size is also important. Figure~\ref{f7} (c) depicts the change of model performance with respect to different batch sizes from 16 to 512. The optimal size is 512. Therefore, we set the batch size to 512 when we train the model. As shown in Figure~\ref{f7} (d), when the number of negative samples is 10, our model achieves the best performance in the Weeplaces dataset. We can conclude that a small number of negative samples can promote the recommendation task, while a bigger one would mislead it.
\section{Related Work}
\subsection{Group Recommendation}
Early work on group recommendation is usually based on the preference integration of group members or the score aggregation of an item across users. The three most common aggregations include the average \cite{DBLP:series/sci/BorattoC11}, the least misery \cite{DBLP:journals/pvldb/Amer-YahiaRCDY09}, and the maximum satisfaction strategy \cite{DBLP:conf/recsys/BaltrunasMR10, DBLP:conf/recsys/BerkovskyF10}. The average method takes the mean score of all members in the group. The least misery is committed to filtering all members' smallest score to please everyone. The maximum satisfaction strategy aims to reach the maximum satisfaction of the member in the group. These three aggregating methods have a significant drawback: they are oversimplified and ignore user interactions in groups.  

Recently, with the successful development of deep learning, models based on the attention mechanism are becoming more and more popular for modeling intra-group user interactions and learning different users' influences \cite{DBLP:conf/sigir/Cao0MAYH18, DBLP:conf/sigir/TranPTLCL19}. For example, GAME \cite{DBLP:conf/sigir/HeCZ20} utilizes the heterogeneous information network and attention mechanism to learn the nodes' multi-view embeddings and members' weights. Yin \emph{et al.} \cite{DBLP:conf/icde/YinW0LYZ19} propose to take the attention mechanism to exploit each user's intra-group user interactions and adopt the bipartite graph embedding to mitigate the data sparsity problem. Although the above methods based on the attention mechanism have made remarkable achievements in group recommendation, they hardly consider the high-order user interactions.

\subsection{Hypergraph-Based Recommendation}
Although the graph neural network approaches have achieved successful results on capturing high-order relations in various tasks \cite{DBLP:conf/iclr/ChenMX18, DBLP:conf/iclr/KipfW17, DBLP:conf/nips/ZouHWJSG19}, these approaches are only appropriate in pairwise connections, which has limitation in expressing complex structures of data. Hypergraph has shown promising potential in modeling complex high-order relations \cite{DBLP:conf/nips/ZhouHS06, DBLP:conf/aaai/TuCWW018, DBLP:conf/iclr/ZhangZ020, DBLP:journals/corr/abs-2001-11181}. Recently, some studies try to combine the hypergraph with the recommender systems to improve their performances. For example, DHCF constructs two hypergraphs for users and items to model their high-order correlations and enhance recommendation models based on collaborative filtering \cite{DBLP:conf/kdd/JiFJZT020}. Xia \emph{et al.} \cite{DBLP:conf/aaai/0013YYWC021} model session-based data as a hypergraph and use the hypergraph neural network to enhance session-based recommendation. Wang \emph{et al.} \cite{DBLP:conf/sigir/WangDH0C20} incorporate the hypergraph into next-item recommendation systems to represent the short-term item correlations. Yu \emph{et al.} \cite{DBLP:conf/www/YuYLWH021} propose a multi-channel hypergraph convolutional network and design multiple motif-induced hypergraphs to exploit high-order user relations patterns. 

Despite the tremendous efforts devoted to these models, they just adopt the paradigm of hypergraph representation learning to model high-order relations. Our models differ significantly from previous approaches in that we use the motif to select informative group interactions.
\vspace{-0.3cm}
\subsection{Self-Supervised Learning in RS}
Self-supervised learning offers a new angle to generate additional supervised signals via transferring the unlabeled data \cite{DBLP:journals/corr/abs-2006-10141, DBLP:journals/corr/abs-2007-08025, DBLP:conf/icassp/BaevskiM20, DBLP:conf/nips/DingZY020,DBLP:conf/kdd/YuY000H21}. Generally, there are two types of self-supervised learning approaches: generative models and contrastive models. Generative models learn to reconstruct the input data and make it similar to raw data \cite{DBLP:journals/corr/abs-2006-08218, DBLP:journals/isci/GaoZYLWX21}. Contrastive models learn discriminative representations by contrasting positive and negative samples \cite{DBLP:conf/cvpr/He0WXG20, DBLP:conf/iccv/GoyalM0M19, DBLP:conf/cvpr/MisraM20, DBLP:conf/naacl/DevlinCLT19}. Recently, some studies attempt to explore the self-supervised learning framework for the recommendation \cite{DBLP:conf/cikm/ZhouWZZWZWW20, DBLP:conf/sigir/XinKAJ20, yao2020self}. For example, Zhou \emph{et al.} \cite{DBLP:conf/cikm/ZhouWZZWZWW20} propose the $S^{3}$-Rec model and design some self-supervised training strategies to pre-train the model. Bert4Rec \cite{DBLP:conf/cikm/SunLWPLOJ19} adopts the Cloze objective to the sequential recommendation by predicting the random masked items in the sequence. Xin \emph{et al.} \cite{DBLP:conf/sigir/XinKAJ20} adopt the cross-entropy as self-supervised Q-learning to fine-tune the next-item recommendation model. Xie \emph{et al.} \cite{xie2020contrastive} propose three data augmentation operations: item crop, item mask, and item reorder, as self-supervision signals to enhance sequential recommendation. 

The closest work to ours is GroupIM \cite{DBLP:conf/sigir/SankarWWZYS20}, which maximizes the mutual information between users and groups to optimize the representation of both. Unlike the above approaches, our work is the first to consider the double-scale node dropout strategies in hypergraph as the supervised signals in group recommendation. We maximize the mutual information between the same hypergraph's different views to enhance data representations and improve recommendation performance.
\vspace{-0.3cm}
\section{Conclusion and Future Work}
In this paper, we proposed a novel neural group recommendation with the hierarchical hypergraph convolutional network and self-supervised learning strategy to capture inter- and intar-group user interactions and alleviate the data sparsity problem. With the triadic motifs, the proposed model can obtain more reliable user interactions beyond groups for more accurate group representations. We innovatively designed a double-scale node dropout strategy as the supervision signals against the data sparsity problem. We conducted extensive experiments on three public datasets to evaluate our models' performance. The experimental results verified the superiority of HHGR and its enhanced version based on self-supervised learning ($S^{2}$-HHGR) compared with other state-of-the-art models. In the future work, we tend to deepen the application of self-supervised learning in group recommendation models and design more general auxiliary tasks for the recommendation to improve the recommendation performance.

\begin{acks}
This work was partially supported by National Natural Science Foundation of China (6217022345), Natural Science Foundation of Chongqing, China (cstc2020jcyj-msxmX0690), Fundamental Research Funds for the Central Universities of Chongqing University (2020CDJ-LHZZ-039),  ARC Discovery Project (DP190101985), and ARC Future Fellowship (FT210100624). 
\end{acks}

\bibliographystyle{ACM-Reference-Format}
\bibliography{sample-base}

\appendix

\end{document}